\documentclass[twocolumn,aps,pra,showpacs,superscriptaddress]{revtex4}

\usepackage{graphicx}
\usepackage{dcolumn}
\usepackage{amsmath}
\usepackage{epsfig}
\usepackage{bm}
\usepackage{amssymb}

\usepackage{hyperref}
\hypersetup{colorlinks=true,linkcolor=blue,citecolor=blue,urlcolor=black}

\newcommand{\ket}[1]{\vert{#1}\rangle}
\newcommand{\bra}[1]{\langle{#1}\vert}
\newcommand{\bkt}[2]{\langle{#1}\vert{#2}\rangle}

\newcommand{\be}{\begin{equation}}
\newcommand{\ee}{\end{equation}}

\begin{document}


\title{Testing Hardy nonlocality proof with genuine energy-time entanglement}


\author{Giuseppe Vallone}
\affiliation{Museo Storico della Fisica e Centro Studi e Ricerche
``Enrico Fermi'', Via Panisperna 89/A, Compendio del Viminale, Roma
I-00184, Italy} 
\affiliation{Dipartimento di Fisica, Universit\`{a} Sapienza di Roma, I-00185 Roma, Italy}

\author{Ilaria Gianani}
\affiliation{Dipartimento di Fisica, Universit\`{a} Sapienza di Roma, I-00185 Roma, Italy}

\author{Enrique B. Inostroza}
\author{Carlos Saavedra }
\affiliation{Center for Optics and Photonics, Universidad de
Concepci\'{o}n, Casilla 4016, Concepci\'{o}n, Chile}
\affiliation{Departamento de F\'{\i}sica, Universidad de
Concepci\'{o}n, Casilla 160-C, Concepci\'{o}n, Chile}

\author{Gustavo Lima}
\email{glima@udec.cl} \affiliation{Center for Optics and Photonics,
Universidad de Concepci\'{o}n, Casilla 4016, Concepci\'{o}n, Chile}
\affiliation{Departamento de F\'{\i}sica, Universidad de
Concepci\'{o}n, Casilla 160-C, Concepci\'{o}n, Chile}

\author{Ad\'an Cabello}
\affiliation{Departamento de F\'{i}sica Aplicada II,
Universidad de Sevilla, E-41012 Sevilla, Spain}

\author{Paolo Mataloni}
\affiliation{Dipartimento di Fisica, Universit\`{a} Sapienza di Roma, I-00185 Roma, Italy}
\affiliation{Istituto Nazionale di Ottica, Consiglio Nazionale delle Ricerche (INO-CNR), L.go E. Fermi 6, I-50125 Firenze, Italy}


\begin{abstract}
We show two experimental realizations of Hardy ladder test of
quantum nonlocality using energy-time correlated photons,
following the scheme proposed by A. Cabello \emph{et al.}
[Phys. Rev. Lett. \textbf{102}, 040401 (2009)]. Unlike,
previous energy-time Bell experiments, these tests require
precise tailored nonmaximally entangled states. One of them is
equivalent to the two-setting two-outcome Bell test requiring a
minimum detection efficiency. The reported experiments are
still affected by the locality and detection loopholes, but are
free of the post-selection loophole of previous energy-time and
time-bin Bell tests.
\end{abstract}



\maketitle


\section{Introduction}


A loophole-free violation of a Bell inequality would prove the
impossibility of describing nature in terms of local hidden
variable theories \cite{bell64phy}, and the possibility of
eternally secure communications \cite{barr05prl}. Among all
versions of Bell's proof, Hardy's \cite{hard92prl, hard93prl}
is probably the simplest. In addition to simplicity, Hardy's
has one interesting feature: it only works for nonmaximally
entangled states, which are precisely the best candidates for a
photonic loophole-free experiment with inefficient detectors
\cite{eber93pra, lars98pra}. To be more specific, the
experimental realization of the two-party, two-setting,
two-outcome Bell test with minimum required detection
efficiency, assuming that all detectors have the same
efficiency \cite{eber93pra, lars98pra}, is equivalent to a test
of Hardy proof.

Standard energy-time and time-bin Bell tests (\emph{e.g.},
\cite{sala08nat}) suffer from a specific loophole called the
post-selection loophole \cite{aert01prl, cabe09prl}, which can
be avoided using a scheme introduced in \cite{cabe09prl}.
Energy-time Bell experiments without the detection loophole and
maximally entangled states have been recently performed using
this scheme \cite{lima10pra}. Moreover, the scheme can be
applied to nonphotonic systems \cite{frus09prb} and can be
extended to multipartite scenarios \cite{vall10pra}.

The aim of this work is to show that energy-time entanglement
can also be used to produce Hardy-type violations of Bell
inequalities free of the post-selection loophole, as a
preliminary step towards a loophole-free Bell test with
photonic random destination sources \cite{scia10ar}. The two
experiments reported in this paper will also show the
feasibility of energy-time entanglement for producing
nonmaximally entangled states which are essential for some
quantum key distribution protocols \cite{cabe00prl}.


\section{Hardy proof}


Hardy proof of nonlocality \cite{hard92prl, hard93prl} can be
summarized as follows. Let us consider two observes, Alice and
Bob, measuring dichotomic (with outputs $-1$ and $1$)
observables. Alice measures $a_0$ and $a_1$, while Bob measures
$b_0$ and $b_1$. Let us define $P(a_i,b_j)$ as the joint
probability of obtaining $a_i=b_j=1$, and $P(\bar a_i,b_j)$ as
the joint probability of obtaining $a_i=-1$ and $b_j=1$. For
any local hidden variable theory with (i) $P(a_0,b_0)=0$, (ii)
$P(\bar a_0,b_1)=0$, and (iii) $P(a_1,\bar b_0=0)$, the
probability $P(a_1,b_1)$ must be equal to zero. However, for
any nonsymmetric pure entangled state, it is always possible to
find observables $a_0$, $a_1$, $b_0$, and $b_1$ such that (i),
(ii), and (iii) are satisfied, while (iv) $P(a_1,b_1)\neq0$
\cite{gold94prl}. $P(a_1,b_1)$ is known as ``Hardy fraction''.
This provides a proof of impossibility of describing quantum
mechanics with local hidden variable theories.

As it was showed by Garuccio and Mermin \cite{merm94ajp,
garu95pra}, Hardy proof can be put in a more generalized
framework writing it in terms of the following inequality:
\begin{equation}
\label{hardyI}
\mathcal S_1\equiv P(a_1,b_1)- P(a_0,b_0)-P(\bar a_0,b_1)- P(a_1,\bar b_0)\leq 0\,,
\end{equation}
which holds for any local hidden variable theory, for any
choice of observables. In this generalized version, there is no
need for vanishing terms in the experimental test, it is only
required that the left hand side of Eq. (\ref {hardyI})
overcomes the sum of terms on the right hand side.
Nevertheless, the interesting feature of Hardy's argumentation
lies in the fact that once one has proven that the
probabilities on the right hand side of the
Garuccio-Mermin-inequality are null, the detection of just one
pair of photon, at the output $a_1=1$ and $b_1=1$ is enough to
refute the local behavior of nature. However, in realistic
conditions, measuring a null probability is not trivial as it
is discussed in Ref \cite{merm94ajp}, and the generalization of
Hardy test to the Bell-type test based on the Clauser-Horne
(CH) inequality \cite{clau74prd} is unavoidable.

The inequality given in Eq. (\ref{hardyI}), is the CH
inequality for an experiment where the following conditions
hold: (a) The quantum efficiency of the detectors is $ \eta =
1$, and (b) photons pairs impinge into the detection
apparatuses. Therefore, Hardy proof can be seen as a special
case of a Bell test based on the CH inequality. This can be
easily demonstrated. Taking into account the above detection
conditions, it comes out that the probabilities of single
photon detections are given by
\begin{subequations}
\begin{align}
P_A(a_i) &=P(a_i,b_j)+P(a_i,\bar b_j)
\\
P_{B}(b_j)&=P(a_i,b_j)+P(\bar a_i, b_j),
\label{marginals}
\end{align}
\end{subequations}
with $i,j=0,1$. For the experimental setup being considered,
the CH inequality can be written as
\be
\begin{aligned}
P(a_1,b_1)+P(a_0,b_1)+ P(a_1, b_0)- P(a_0,b_0)
\\
-P_A(a_1) -P_B(b_1) \leq0, \label{chshar}
\end{aligned}
\ee which turns to be Eq.~(\ref{hardyI}), when one replaces the
marginal probabilities (\ref{marginals}) into
Eq.~(\ref{chshar}). It is also worth to mention that, under
these conditions, the CH inequality is equivalent to the
Clauser-Horne-Shimony-Holt (CHSH) inequality \cite{clau69prl}.
Therefore, Hardy proof can also be seen as a particular case of
the nonlocality tests based on the CHSH inequality. Indeed, any
experimental setup prepared for testing the CHSH inequality,
can also be used to test Hardy proof if the degree of
entanglement of the state being generated can be manipulated.

\begin{figure}[t]
\includegraphics[width=8.5cm]{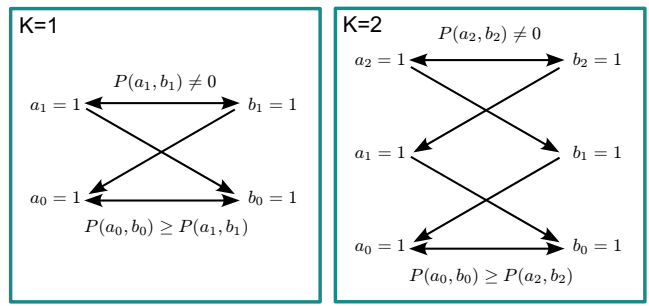}
\caption{Ladder proof schemes for $K=1$ and $K=2$.}
\label{Fig:scheme}
\end{figure}

Hardy proof can be generalized by considering a system in
which, having defined $K+1$ dichotomic observables $a_k$ and
$b_k$ ($k=0,\ldots,K$), the following probabilities hold:
\begin{equation}
\label{Prob}
\begin{aligned}
&P(a_K,b_K) \neq 0\,,
\\
&\left.
\begin{aligned}
&P(\bar a_{k-1},b_k) = 0\,,
\\
&P(\bar a_k,b_{k-1}) = 0\,.
\\
\end{aligned}
\right\}k=1,\ldots,K
\\
\end{aligned}
\end{equation}
When $K$ is larger than 1, the test can be interpreted as a
chained violation, and can be represented as a ladder
\cite{bosc97prl, barb05pla} on which each step implies the one
below (see Fig. \ref{Fig:scheme}).


Let us consider, for example, $K=2$. If the first equation in
\eqref{Prob} holds, then there exists a nonzero probability
that both $a_2=1$ and $b_2=1$ occur. The second and third
equations in \eqref{Prob} state that the probabilities of
$a_2=1$ and $b_1=-1$, or $a_1=-1$ and $b_2=1$ are zero. In this
case, $a_1=1$ and $b_1=1$ should have been observed. The same
applies to the lower step, reaching in this way the bottom of
the ladder. At this point, we obtain that both $a_0=1$ and
$b_0=1$ should have been measured, and thus the probability
$P(a_0,b_0)$ should be different from zero. In local theories
$P(a_0,b_0)$ should be at least equal to $P(a_K=1,b_K=1)$. If
there exist a system in which this probability is vanishing, a
classical theory would not be able to describe the system, and
the Hardy inequality would be violated. A system like that can
be implemented by the setup shown in the next Section. It can
be tested by generalizing Eq. \eqref{hardyI} for the ladder
proof case:
\begin{equation}
\label{inequality}
\begin{aligned}
\mathcal S_K\equiv &P(a_K,b_K)-P(a_0,b_0)
\\
&-\sum^K_{k=1}\left[P(a_k,\bar b_{k-1})+P(\bar a_{k-1},b_k)\right]\leq0\,.
\end{aligned}
\end{equation}


\section{Experiment}


\subsection{Energy-time Hardy test}

\begin{figure}[t]
\includegraphics[width=0.5\textwidth]{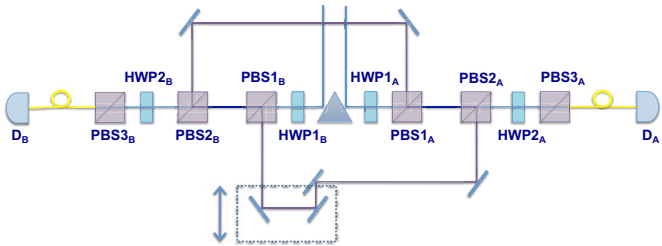}
\caption{Experimental setup. A step-by-step translation stage allows us to create
the indistinguishability condition ($L_A - S_A = L_B - S_B$).
Coincidence windows is $3 ns$ and, due to continuous wave pumping,
the probability of two photon pair events is negligible.
Interference filters select a bandwidths of $3.5$~nm.
The radiation is coupled into single-mode optical fibers and sent to
pigtailed avalanche photo-counting modules connected to a
circuit used to record the single and the coincidence counts.
The phase of the interferometer can be controlled using a piezoelectric
stage on which PBS2$_A$ is assembled.} 
\label{Fig:setup}
\end{figure}

A Hardy test can be, in principle, implemented by using any
entangled state, except the one which is maximally entangled.
Our capacity to generate two photons correlated in the
energy-time degree of freedom in partially entangled states,
and the ability to detect them with controllable
interferometric techniques allows for implementing a Hardy test
with the experimental setup of Fig.~\ref{Fig:setup}. Let us
consider the energy-time state of two down-converted photons
$\ket{\Phi}\equiv \alpha \ket{S_A S_B}+ \beta\ket{L_A L_B}$ and
define the following $K+1$ spatial measurement basis, in each
direction $A_k$, $B_k$, where $k=0,\ldots,K$:
\begin{subequations}
\begin{align}
\label{base1} \ket{A_k}=\cos\theta_k\ket{S}+\sin\theta_k\ket{L},
\\
\label{base1orto}
\ket{A^\bot_k}=\sin\theta_k\ket{S}-\cos\theta_k\ket{L},
\\
\label{base2} \ket{B_k}=\cos\theta_k\ket{S}+\sin\theta_k\ket{L},
\\
\label{base2orto}
\ket{B^\bot_k}=\sin\theta_k\ket{S}-\cos\theta_k\ket{L}.
\end{align}
\end{subequations}
Let us define the operators $a_k$ ($b_k$ is defined similarly)
having outcome 1 or $-1$ when the state $\ket{A_k}$ or
$\ket{A^\bot_k}$ is respectively detected. In order to prove
nonlocality we need to satisfy the conditions written in
\eqref{Prob}. That is,
\begin{subequations}
\label{P1}
\begin{align}
&P(a_K,b_K) = | \bra{A_K}\bkt{B_K}{\Phi}|^2 \neq 0,
\\
&P(\bar a_{k-1},b_k) = | \bra{A_{k-1}^{\bot}}\bkt{B_k}{\Phi}|^2 = 0,
\\
&P(a_k,\bar b_{k-1}) = | \bra{A_k}\bkt{B^\bot_{k-1}}{\Phi}|^2 = 0\,.
\end{align}
\end{subequations}
Moreover, we need that the following condition holds:
\begin{equation}
 \label{P4}
P(a_0,b_0) = | \bra{A_{k-1}^{\bot}}\bkt{B_{k-1}^{\bot}}{\Phi}|^2 = 0,
\end{equation}
The values of $\theta_k$ solving the previous equations are given by the relations:
\begin{equation}\label{angolo}
\sin\theta_k=\left(-1\right)^k\frac{T^{k+\frac{1}{2}}}{\sqrt{T^{2k+1}+1}},
\end{equation}
with $t=\alpha/\beta$ related to the degree of entanglement.
The Hardy fraction $P(a_K,b_K)$ is then given by
\begin{equation}
\label{P1teo}
 P(a_K,b_K)=\frac{t^2\left(t^{2K}-1\right)^2}{\left(t^{2K+1}+1\right)^2\left(1+t^2\right)}
 \end{equation}
\begin{figure}[t]
\includegraphics[width=0.5\textwidth]{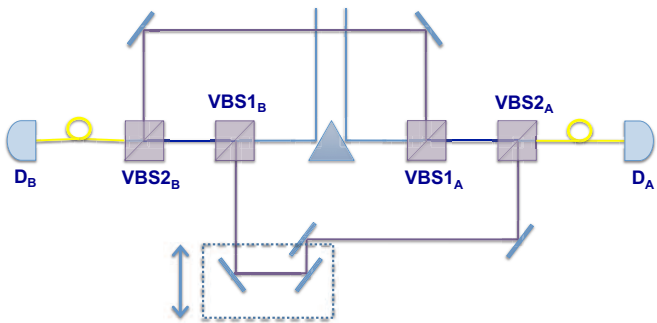}
\caption{scheme needed to implement a Hardy test using energy-time entanglement.} 
\label{Fig:setup1}
\end{figure}

When $K=1$ ($K=2$), this function is maximized at $t=t^*_1\simeq0.46$ ($t=t^*_2\simeq0.57$) with value
$P(a_1, b_1)_{max}\simeq0.09$ ($P(a_2, b_2)_{max}\simeq0.17$). 
When $K=1$, only $9\%$ of particles
violates locality, but this fraction can be amplified using a
higher value of $K$. It has been shown in fact that when
$K\longrightarrow\infty$, $P(a_K, b_K)_{max}\longrightarrow
50\%$ \cite{bosc97prl}.


\subsection{Experimental setup}

\begin{figure}[t]
\includegraphics[width=0.50\textwidth]{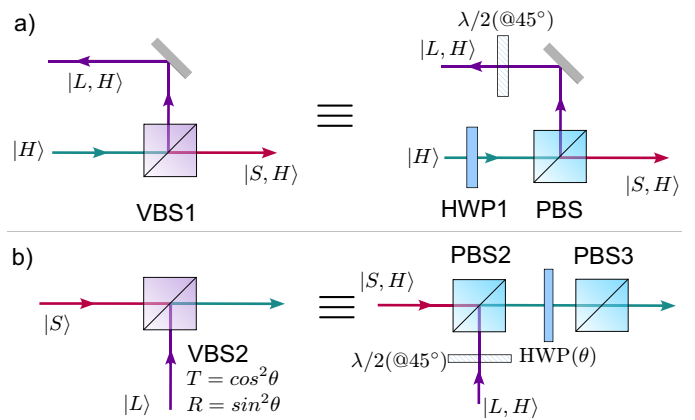}
\caption{Generation of nonmaximally entangled state (a) and projection on the measurement basis (b).
This setup is repeated on both modes A and B, and is used to generate a nonmaximally entangled state
and project on the desired base, as described in the text.} \label{Fig:stateprep}
\end{figure}

We generated energy-time correlated photons by spontaneous
parametric down-conversion (SPDC) \cite{burn70prl, hong87prl}.
A 1 mm $\beta$-barium borate crystal (BBO) shined by an UV
laser beam generated pairs of photons at wavelength $532$ nm.
The emission time of each pair is unpredictable due to the long
coherence length ($\geq 1$ m) of the pump laser beam. The two
photons generated with horizontal polarization are sent through
two unbalanced interferometers as shown in
Fig.~\ref{Fig:setup}. As it will be discussed shortly this is a
modified version of the interferometric scheme previously used
by us \cite{lima10pra}. As for the setup previously used, the
geometry of these interferometers has been showed to allow for
more genuine tests of quantum nonlocality with energy-time
correlated photons \cite{cabe09prl}. In this case, even though
the experiment is still constrained by the locality and
detection loopholes \cite{sant04fp}, it is not necessary to
assume any other auxiliary assumption for validating it as a
conclusively Bell test \cite{bell64phy}. These interferometers
are unbalanced and so one can refer to their arms as short
($S$) and long ($L$). The optical paths followed by the
down-converted photons are such that coincidences between
detectors $D_{A}$ and $D_{B}$ are measured only when they both
propagate through the short or long photon paths.

In order to perform a Hardy test, nonmaximally entangled states
are needed, and it is also possible to use just two detectors
instead of the four used in the previous experiment
\cite{lima10pra}. The new scheme is shown in Fig.
\ref{Fig:setup1}: variable beam splitters ($VBS1_{A,B}$) are
used on both modes in order to prepare the non maximally
entangled state:
\begin{equation}
 \ket{\psi} = \alpha\ket{S_AS_B}+\beta\ket{L_AL_B}.
\end{equation}
In this way, the ratio $t=\alpha/\beta$ can be controlled by
using the following relation between the transmittivities
($T_A$ and $T_B$) and reflectivities ($R_A$ and $R_B$) of
$VBS1_{A,B}$ and $t$:
\begin{equation}
 \sqrt{\frac{T_AT_B}{R_AR_B}}= \frac{\alpha}{\beta}=t.
\end{equation}
The two $VBS2_{A,B}$ can be used to project into the states of
Eq. \eqref{base1}-\eqref{base2orto}, providing that their
transmittivities and reflectivities are linked to $\theta_k$
being $\sqrt{R}=\sin{\theta_k}$ and $\sqrt{T}=\cos{\theta_k}$.

In Fig. \ref{Fig:setup} we show the setup we actually used for
the experiment. It is equivalent to the one shown in Fig.
\ref{Fig:setup1} but it doesn't use variable beam splitters.

The VBSs used to prepare the state, namely $VBS1_{A,B}$, are
implemented in the way shown by the scheme of Fig.
\ref{Fig:stateprep}a): the polarization of both photons $A$ and
$B$ is changed by a half wave plate (HWP1) and then by a
polarizing beam splitter (PBS1) each photon is splitted in the
long and short paths. After the PBS1 we will have on each mode,
in transmission, the state: $\ket{S}\ket{H}$, while on
reflection, the state: $\ket{L}\ket{V}$. By rotating HWP1, one
can change the amount of light being reflected and transmitted,
allowing to create a nonmaximally entangled state.

To implement the two VBS2, we used the scheme reported in Fig.
\ref{Fig:stateprep}b): since the $\ket{S}$ ($\ket{L}$) mode is
horizontally (vertically) polarized, any state encoded into the
energy-time degree of freedom is converted into polarization
encoding by the PBS2. Then the projection on the desired state
can be implemented by standard polarization analyzer (HWP2
rotated at $\theta_k/2$ and PBS3).


\subsection{Experimental results}

\begin{table}[t]
\caption{\label{table}Experimental probabilities needed to violate the inequality $\mathcal S_K\leq0$
for $K=1$ and $K=2$. The reported data are obtained with the value of $t$ that maximize the violation, namely 
$t=t^*_1\simeq0.46$ for $K=1$ and $t=t^*_2\simeq0.57$ for $K=2$.}
\begin{ruledtabular}
\begin{tabular}{cccc}
\multicolumn{2}{c}{$K=1$, $t=t^*_1$} & \multicolumn{2}{c}{$K=2$, $t=t^*_2$}
\\
\hline
$P(a_1,b_1)$&$0.095\pm0.005$
&$P(a_2,b_2)$&$0.170\pm0.008$
\\
$P(a_1,\bar b_0)$&$0.005\pm0.001$
&$P(a_2,\bar b_1)$&$0.007\pm0.002$
\\
$P(\bar a_0,b_1)$&$0.005\pm0.001$
&
$P(\bar a_1,b_2)$&$0.009\pm0.002$
\\
$P(a_0,b_0)$&$0.007\pm0.001$
&
$P(a_1,\bar b_0)$&$0.009\pm0.002$
\\
$\mathcal S_1$&$0.078\pm0.005$
&$P(\bar a_0,b_1)$&$0.009\pm0.002$
\\
&&$P(a_0,b_0)$&$0.011\pm0.002$
\\
&&$\mathcal S_2$&$0.124\pm0.009$
\end{tabular}
\end{ruledtabular}
\end{table}

Our experiment aimed at the violation of the Hardy inequality
\eqref{inequality} for $K=1$ and $K=2$. For this purpose, we
measured the probabilities described by equations \eqref{P1}
and \eqref{P4} as:
\begin{equation}
P=\frac{C\left(a_i=\alpha, b_j=\beta\right)}{C_{TOT}},
\end{equation}
where $i,j$ are the directions required, and
$\alpha,\beta=-1,1$ as previously specified.
$C\left(a_i=\alpha, b_j=\beta\right)$ is the number of
coincidences obtained measuring on the projected state needed
for both modes, while $C_{TOT}= C\left(H,H\right) +
C\left(V,V\right) + C\left(H,V\right) + C\left(V,H\right)$ is
the sum of the number of coincidences over all the possible
outcomes in the base $\{\ket H,\ket V\}$.


\begin{figure}[t]
\includegraphics[width=8cm]{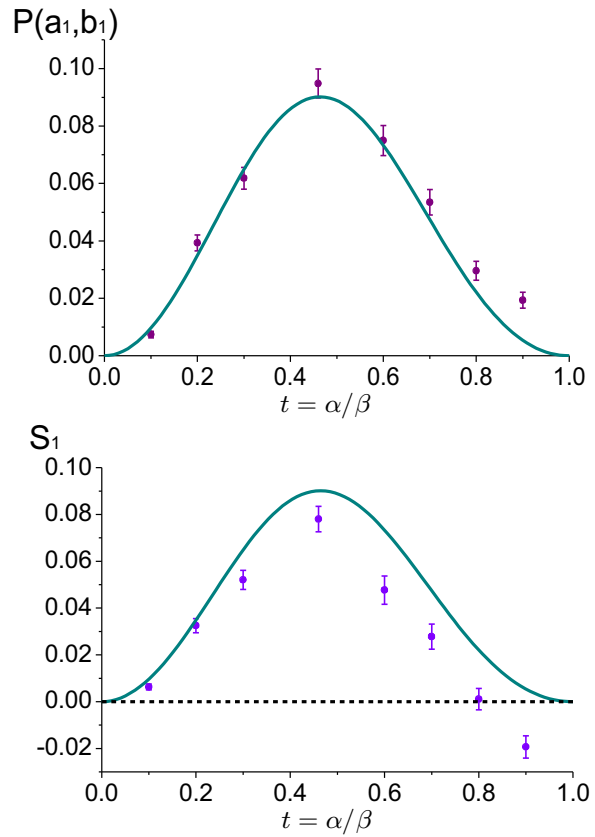}
\caption{Experimental results for K=1.
This two graphs show the result obtained for the one-step ladder proof.
The first one shows the agreement between the data and the theoretical prediction (continuous line) for the probability $P_(a_1,b_1)$.
The second one shows the obtained values of $\mathcal S_1$ compared to the theory (continuous line). 
The dotted line represents the classical bound: 
when $\mathcal S_1$ is greater than 0
the inequality \eqref{hardyI} is violated.} \label{Fig:graph1}
\end{figure}


\begin{figure}[t]
\includegraphics[width=8cm]{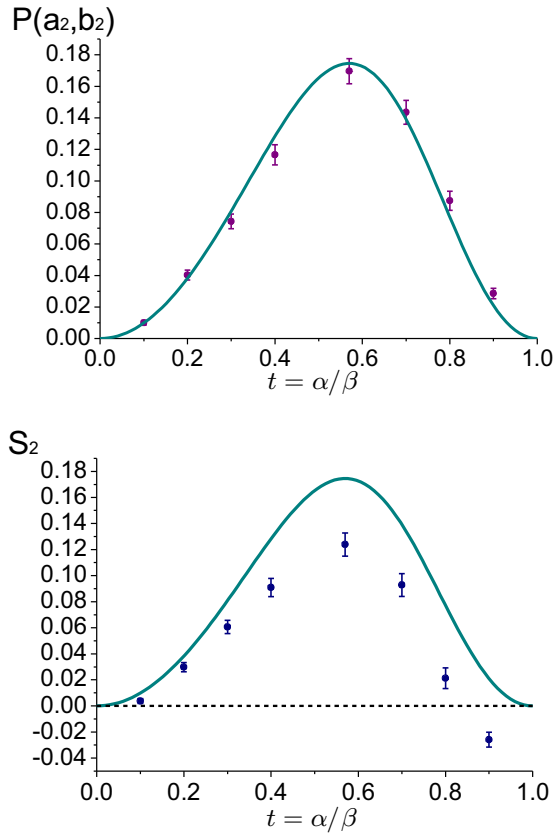}
\caption{Experimental results for K=2. The results obtained for a two-steps ladder proof are shown.
As in the previous figure, in the first one is reported the probability $P_(a_2,b_2)$ along with the theoretical function,
while in the second there is the inequality violation. When $\mathcal S_2$ is greater than 0 
the inequality \eqref{inequality} is violated.} 
\label{Fig:graph2}
\end{figure}


In table \ref{table} we show the experimental probabilities obtained for $K=1$ and $K=2$
when $t=t^*_1$ and $t=t^*_2$ respectively.
In Figs. \ref{Fig:graph1} and \ref{Fig:graph2} we show the
probabilities obtained for $K=1$ and $K=2$ for different values of $t=\alpha/\beta$. 
Each figure shows the graph of the experimental $P(a_K,b_K)$ compared to the theoretical values described by
equation \eqref{P1teo} and the graph of the inequality
violation $\mathcal S_K$ defined in \eqref{hardyI} and \eqref{inequality}. 
The obtained values of $P_(a_K,b_K)$ are in good agreement with the theoretical predictions for both
$K$ values. The inequality is not violated for large values of
$t$, $t \geq 0.8$ (see Fig. \ref{Fig:graph1} and
\ref{Fig:graph2}). This can be due to the imperfect
experimental visibility. In fact, we measured $V\sim96\%$ when
the state is maximally entangled ($t=1$). However, this value
is not enough to allow the probabilities $P(a_k,\bar b_{k-1})$, 
$P(\bar a_{k-1},b_k)$ and $P(a_0,b_0)$ to vanish completely. This feature is
more evident when the degree of entanglement increases since,
in this case, the $\ket{SS}$/$\ket{LL}$ interference is more
important.


\section{Conclusions}


Recently introduced schemes for energy-time and time-bin
entanglement can be improved for a Bell test with nonmaximally
entangled states free of the postselection loophole. The
important point is that these tests are less demanding in terms
of detection efficiency than those based on maximally entangled
states (which were the states used in previous Bell tests with
energy-time and time-bin entanglement), even when the
postselection is taken into account \cite{scia10ar}. The
experiments reported in this paper still suffer the detection
and nonlocality loopholes, but show the feasibility of
energy-time Bell tests with nonmaximally entangled states and
free of the postselection loophole.

In addition, the ability for projecting energy-time correlated
photons in this more general set of projections may also have
other important applications. It could be useful for
entanglement witnesses, quantum tomography, and for some
cryptographic protocols requiring nonmaximally entangled
states.


\begin{acknowledgments}
G.L. thanks to grant FONDECYT 11085055 and PBCT-PDA21. A. C.
acknowledges support from Spanish MCI Project FIS2008-05596 and
the Wenner-Gren Foundation. C.S. acknowledges FONDECyT 1080383.
\end{acknowledgments}


%

\end{document}